# Unveiling City Jam-prints of Urban Traffic based on Jam Patterns

Guanwen Zeng[1,2], Nimrod Serok[3], Efrat Blumenfeld Lieberthal[3], Jinxiao Duan[4], Shiyan Liu[2], Shaobo Sui[2], Daqing Li[2], Shlomo Havlin[1,*]

1. Department of Physics, Bar-Ilan University, Ramat Gan 52900, Israel.
2. School of Reliability and Systems Engineering, Beihang University, Beijing 100191, China.
3. Azrieli School of Architecture, Tel Aviv University, Tel Aviv 6997801, Israel.
4. School of Economics and Management, Beihang University, Beijing 100191, China.

**Email**
*Correspondence should be addressed to: havlins@gmail.com

**Abstract**
We analyze the patterns of traffic jams in urban networks of five large cities and an urban agglomeration region in China using real data based on a recently developed jam tree model. This model focuses on the way traffic jams spread through a network of streets, where the first street that becomes congested represents the bottleneck of the jam. We extended the model by integrating additional realistic jam components into the model and find that, while the locations of traffic jams can vary significantly from day to day and hour to hour, the daily distribution of the costs associated with these jams follows a consistent pattern, i.e., a power law with similar exponents. This distribution pattern appears to hold not only for a given region on different days, but also for the same hours on different days. This daily pattern of exponent values for traffic jams can be used as a fingerprint for urban traffic, i.e., jam-prints. Our findings are useful for quantifying the reliability of urban traffic system, and for improving traffic management and control.

**Introduction**
Traffic jams have been a long-term worldwide painful problem for travelers and urban traffic administrators. They are often considered as the result of the decision-making processes of individual travelers (i.e., agents) who display bounded rationality in a system with limited transportation resources. The complexity of traffic jam is expected to significantly increase as urban populations and the size of cities grow. This is particularly true in megacities and urban agglomerations, where traffic jams can be challenging to address. With the emergence of new technologies, i.e., Internet of Things, cloud computing, big data, and artificial intelligence, modern transportation systems could be improved and become more intelligent and efficient, known as the Intelligent Transportation Systems (i.e., ITS). The ITS integrates various types of networks, including cyber networks (i.e., sensing and communication networks), physical networks (i.e., traffic network) and social networks (i.e., traveling demand and traveling trajectory networks). On one hand, ITS is aiming at reducing traffic jams by using comprehensive sensing, communicating, and monitoring technologies. On the other hand, the introduction of these new technologies could also enhance the interdependent interactions among different elements in the transportation system (i.e., travelers, cars, roads, devices, etc.), potentially creating new risks. For instance, it has been found that as the strength of these interdependent interactions grows, cascading failures appear and lead to abrupt collapses of the system of systems even from a small region of initial failures[1,2].

A key challenge in addressing traffic congestion is detecting and mitigating bottlenecks, i.e., jam sources, in urban traffic. If traffic bottlenecks are not alleviated at their early stages, they may evolve to regional congestion or even gridlocks[3,4] in the traffic network. Most existing studies have focused on the cause of traffic bottlenecks from a road-level perspective[5-9], which have laid solid foundation for research on traffic congestion mechanism. Meanwhile, many related solutions have also been suggested and applied to bottleneck mitigation, such as diversion and lane restriction[10], signal timing[11], speed restriction[12], and connected vehicles system[13,14]. Besides road-level perspective, some scholars have also sought for the emergence of network-level bottlenecks. These studies often focused on uncovering the key elements that influence the spreading process in the network[15] to identify the potential bottlenecks. In recent years, studies based on percolation theory have also attempted to uncover the potential bottlenecks of urban traffic network by considering their effects on the global traffic organization in the city. For example, Li et al. manifested that bottlenecks in different locations can play different roles in the organization process of traffic networks, and addressing these network-level bottlenecks can significantly improve global traffic[16]; Hamedmoghadam et al. uncovered traffic bottlenecks by designing a percolation-based framework based on flow heterogeneity of the traffic network[17]. However, the practical mitigation of traffic bottlenecks is still complex due to the complexity behaviors of the travelers[18,19] and cascading jams in traffic networks[20].

Since it can be important yet difficult to control network bottlenecks, researchers have also analyzed traffic jams based on their global patterns. Geroliminis and Daganzo[21] found a fundamental relationship called Macroscopic Fundamental Diagram (MFD) in global traffic (i.e., the relationship between average flow, average density, and average speed of all roads) in a region where the distribution of cars is relatively homogeneous. Based on MFD, many corresponding traffic control strategies, e.g., perimeter control, have been designed[22-25]. Zeng et al. found that traffic dynamics could represent multiple network percolating states based on percolation theory[26,27]. In addition, another study focused on the recovery behavior of traffic jams, which proposed a measure for evaluating the spatio-temporal resilience of urban traffic congestion[28]. They discovered a scaling law for the distribution of resilience, suggesting that there is an inherent universal behavior behind traffic resilience which is independent of details at the microscopic scale. A recent study of Serok et al. developed a method for identifying "traffic jam trees" based on the spatio-temporal relationships between the sources of the congestion (defined as bottlenecks) and the streets it affects[29]. While the study analyzed the overall daily impact (or "jam cost") of traffic congestion in specific cities, the evolution of traffic congestion during the day is unclear yet critical for network jam formation.

In the present study, we extend the jam tree model of Serok et al.[29] and study congestion patterns in urban traffic of five different cities and an urban agglomeration region in China. We analyzed the evolution of jam tree cost (in vehicle hours), and particularly the characteristics of jam trees in different scenarios (i.e., rush hours and non-rush hours). We find that although the locations of the jam trees vary significantly from day to day, the distribution of their costs follow a similar pattern every day, i.e., a power law distribution with similar exponents. We also find that the exponents for the same city in different days are very similar not only for the entire day, but also for the same hours in different days. On the other hand, they are different for different cities. This suggests that the daily pattern of jam tree exponents can be used as a fingerprint of the city, i.e., as a unique characteristic of a city's traffic during a 24-hour period. Our results suggest a new way to classify and compare the evolution of traffic in different cities, which can be useful for evaluating traffic reliability as well as for strategies for improving traffic management and control by observing the changes in the fingerprint.

**Results**

The traffic jam tree model[29] relates the propagation of traffic congestion to the growth of a tree: the downstream road that becomes congested first represents the trunk of a tree, while the upstream roads that become later congested represent the branches which sprout from the trunk. Here, we show based on the tree-like structure, the characteristics of the emergence, growth, and dissipation of the urban traffic congestion. By analyzing the jam trees, one can identify the trunks of trees at a given time (see Fig. 1). These can be regarded as the bottlenecks representing the origins of large-scale urban congestion.

In this paper, we extend the definition of jam trees to include additional types of jam trees and analyze their statistical properties using big real data of high resolution (i.e., 1 minute). Specifically, we analyze here the traffic patterns of typical regions in China over the course of a full day. Our generalization of the jam tree model enables us to identify additional types of jam trees beyond those identified in Ref. [29], and to determine their jam trunks, as well as their corresponding duration, sizes and costs. To identify the jam trees in urban traffic, our first step is to calculate the initial time and duration of the congestion (until a given time) for each road segment in the urban traffic network. The duration of the congestion is defined as the successive time that a road segment has been congested (i.e., the relative velocity of the road is below a predefined threshold[16], see also Methods). The basic assumption is that the traffic congestion is initiated on a downstream road(s) and propagates over time to its neighboring upstream roads and then to its next nearest neighbors and so on. Therefore, we can identify and follow the evolution of each jam-tree in the traffic network by analyzing the congestion on different road segments and the connectivity of the roads. The methodology of identifying the jam trees in urban traffic is in principle based on Ref. [29]. However, in this work, we extend the definition of a jam-tree to include further general cases; in particular, cases where several jam trees may overlap and share the same trunk or branches. The core idea of our extended approach is to associate the influence of a jam to its potential trunk(s) in a reasonable way (see Fig. 1), that is, to share the influences from parallel jammed downstream roads or to determine the original influence from successive downstream roads. This approach is highly necessary as these cases are common, particularly in megacities with complex traffic patterns; however, they have not been discussed in the previous study. Further details are given in the Methods section.

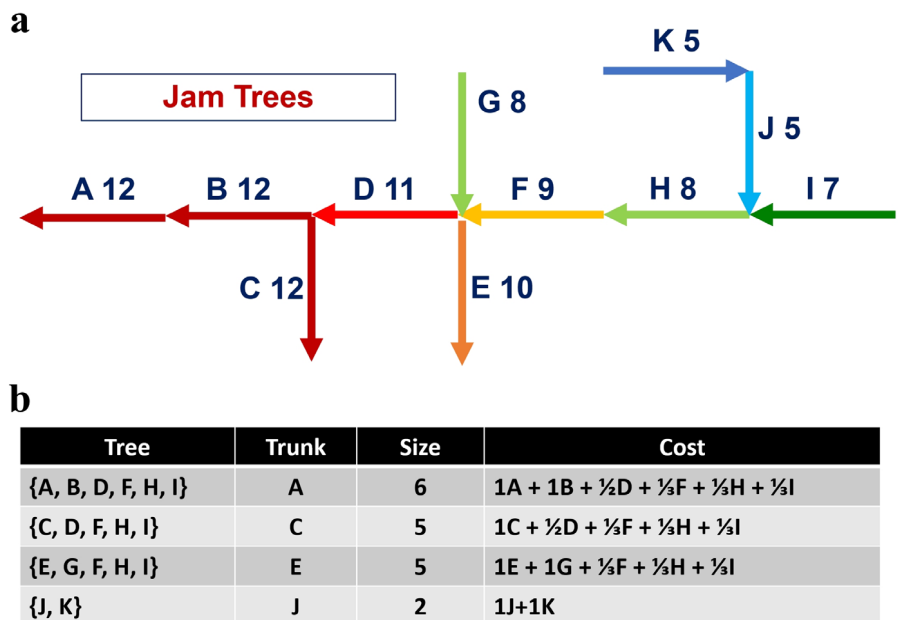

**Fig. 1 Definition of a jam tree. (a)** Examples of several jam trees. Each directed link represents a road segment, with an arrow indicating the direction of traffic flow on it. The number above each link indicates its jam duration, i.e., number of successive time intervals of traffic jam. Here, each time interval represents 10 minutes. If the difference in jam duration between two adjacent segments is less than a threshold, (in this study 2 intervals, i.e., 20 minutes), they are considered part of the same "jam tree"[29]. This is based on the assumption that there is a causal relationship between upstream and downstream traffic flow both in space and time. **(b)** Key information about the presented jam trees in **(a)**. The table presents information of each jam tree, including its trunk, size, and cost. Each jam tree has only a single specific trunk based on the above definition. The size of a jam tree is determined by the number of segments that belong to it, including both the main trunk and associated branches, while the temporal cost of each jam tree is defined as the weighted sum cost of its trunk and branches (See Methods).

As can be seen in Fig.1, each letter (A to K) in the figure represents a road segment, with an arrow pointing in the direction of the traffic flow on that segment. The number above each link is the jam past duration (in time interval units) of the link at a given time. The table in Fig. 1b demonstrates how we calculate the tree size and cost. The size of a tree is defined as the number of road segments that this jam tree contains, while its cost is defined as the weighted sum cost of its related road segments (the cost of a single road segment is calculated by following Ref. 29, see also Methods). However, in this work, we extend the methodology to encompass additional scenarios involving the aggregation of segment costs linked to a specific tree. Such scenarios were not addressed in the previous model. If a road segment can be associated to multiple trunks, e.g., link F in Fig. 1b, which is a shared branch of trunks A, C, and E, the cost of this segment is divided equally among all the trunks that the link can be associated with. In this case, 1/3 of the cost would be assigned to each trunk. This is because each trunk has a priori an equal chance to induce this link congestion. Based on this definition, we identify the jam trees in two typical megacities in China (Beijing and Shenzhen) on different days. We summarize the daily number of the jam trees (trunks) according to their sizes in Table 1. In the following, we present results for additional three large cities as well as an urban agglomeration region.

**Table 1: The number of trunks (trees) according to their sizes (daily average)**

| Size threshold | **Beijing** (52968 links) | **Shenzhen** (22248 links) |
|---|---|---|
| $\geq 1$ | 34647 (65.4%) | 13341 (60.0%) |
| $\geq 5$ | 15483 (29.2%) | 6371 (28.6%) |
| $\geq 10$ | 3850 (7.3%) | 1905 (8.6%) |
| $\geq 20$ | 416 (0.8%) | 218 (1.0%) |

**Note**: The percentage in the parenthesis is derived considering the number of trunks out of the number of total links in each city, i.e., 52968 for Beijing and 22248 for Shenzhen (e.g., 65.4%=34647/52968). These percentages represent the relative jam scale of a city during the day.

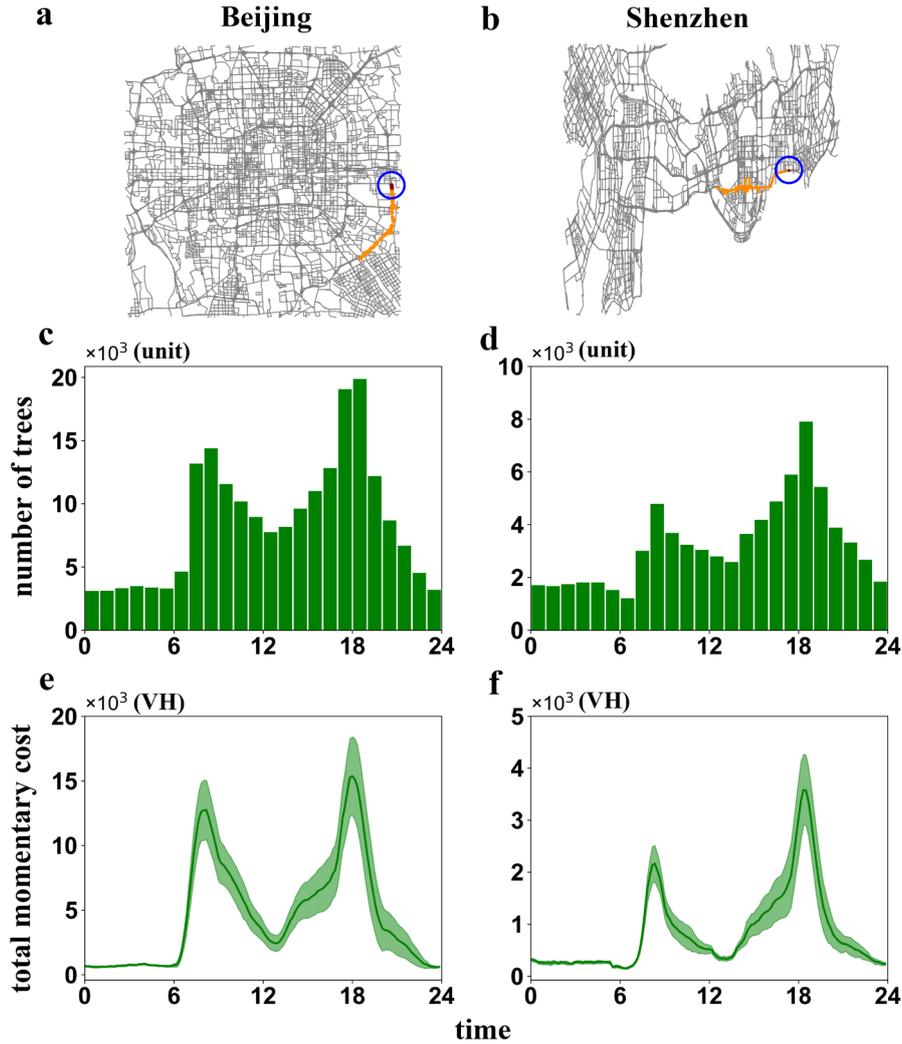

**Fig. 2 Temporal evolution of jam trees during workdays.** Demonstration of a typical jam tree with large influence during the day in October 26 in **(a)** Beijing and **(b)** Shenzhen, where the dark red link represents the trunk (surrounded by blue circle) while the orange links represent the branches. The number of existing jam trees during the day (average on 17 working days with the resolution of 1 hour) in **(c)** Beijing and **(d)** Shenzhen. The total momentary cost of jam trees during the day (with the resolution of 10 minutes) in **(e)** Beijing and **(f)** Shenzhen. The error-bars (lighter green) which are surprisingly small are the variances of the total momentary cost in 17 working days.

During its lifespan, a jam tree will grow and then dissipate. It is important to understand for each time during the day the information about jam trees, each connected to a specific trunk, i.e., what is their impact on the traffic system. In Fig. 2a and Fig. 2b, we demonstrate a typical jam tree of large influence during the day of October 26 in Beijing and Shenzhen, respectively. Next, to quantitively address this issue, we begin by analyzing the evolving number of jam trees throughout the day. By calculating the number of trunks (each represents a jam tree) during the day, we find two distinct peaks: one between 7:00-9:00 and the other between 17:00-19:00 in both Beijing and Shenzhen, which represent the rush-hour windows in both cities (Fig. 2c and Fig. 2d).

Note that, in both cities, we find more jam trees during the afternoon rush hours compared to the morning rush hours. This is probably because in the afternoon, individuals are heading to a wider variety of locations (such as their homes), as opposed to the morning when the travel is primarily directed towards central work places. The number of jam trees in each time unit (Fig. 2c and 2d) depicts the *frequency* of existing jam trees but it does not give us information about their actual *importance* (i.e., size or cost). Thus, next we calculate the total momentary cost by summing up the momentary cost of all jam trees in the urban traffic network at a given time, and analyze how this value evolves during the day. To further evaluate the influence of different traffic jams, particularly in terms of the economic or social costs imposed by the trunk of a jam tree, we calculate the vehicle hours (VH) spent in traffic on the road segments of the jam tree. This measure, known as the cost of the jam tree, is calculated using the methodologies outlined in Ref. 29. The cost of a link represents the total extra travel time taken by drivers to cross the road segment above the optimal travel time under normal traffic conditions. Thus, the cost of a link at a given time $t$, $C(t)$, can be calculated by the equation[29]:

$$C(t) = d * \left( \frac{1}{v(t)} - \frac{1}{v_{op}} \right) * \frac{q(t) * l}{\frac{60}{T}}, \qquad (1)$$

where $d$ is the length of the link, $v(t)$ is the current velocity on the link at time $t$, $v_{op}$ is the optimal velocity on the link which can be calculated by the macroscopic fundamental diagram, $q(t)$ is the current flow on the link at time $t$, $l$ is the number of lanes in the link, $T$ is the time interval we set as before (i.e., 10 minutes). Note that in our dataset we do not have information on the number of lanes in each link. Therefore, we estimate this value based on the rank of each link (details can be found in Methods). According to Eq. (1), it is necessary to initially evaluate traffic flow on each link. The flow $q(t)$ on the link can be calculated by:

$$q(t) = v(t) * k(t), \qquad (2)$$

where $k(t)$ is the current car density on the link in time $t$. Besides, the relationship between the velocity and density on the link is based on the generalized car-following model assumed as[30]:

$$\left( \frac{v(t)}{v_f} \right)^{1-m} = 1 - \left( \frac{k(t)}{k_j} \right)^{l-1}, \qquad (3)$$

where $v_f$ is the free-flow velocity on the link we assume to be the velocity limit on the link, $k_j$ is the jam density on the link which is set as 150 veh/km. The values of the parameters $m$ and $l$ are used to determine the shape of speed-density relation curve, which are usually derived from empirical data and yield $m=0.8$ and $l=2.8$ in this case. Based on the above equations, we can derive the flow on each link at any specific time with our velocity dataset.

As seen in Fig. 2e and Fig. 2f, the evolution of the total momentary cost also presents, for both cities, two peaks during rush hours. These results not only indicate, as expected, that during rush hours the urban transportation network has significantly more and larger traffic jams, but also quantify their momentary cost. Moreover, when comparing Beijing and Shenzhen, we find that the total momentary cost during rush hours in Beijing is approximately 4-5 times as that in Shenzhen, which is most probably because Beijing is a larger city. It can be also seen that in both cities the evening rush hour is more congested than the morning rush hour, with a higher total momentary cost during the evening rush hours of over 15000 VH in Beijing and over 3000 VH in Shenzhen. However, it is worth noting, as seen in Fig. S1 in the *Supplementary Information*, that when it comes to the average momentary cost at each moment, the situation can be rather different. In Shenzhen, the average cost during the two rush-hour periods is almost the same, while in Beijing the average momentary cost during the morning rush hours is larger than that during the evening rush hours.

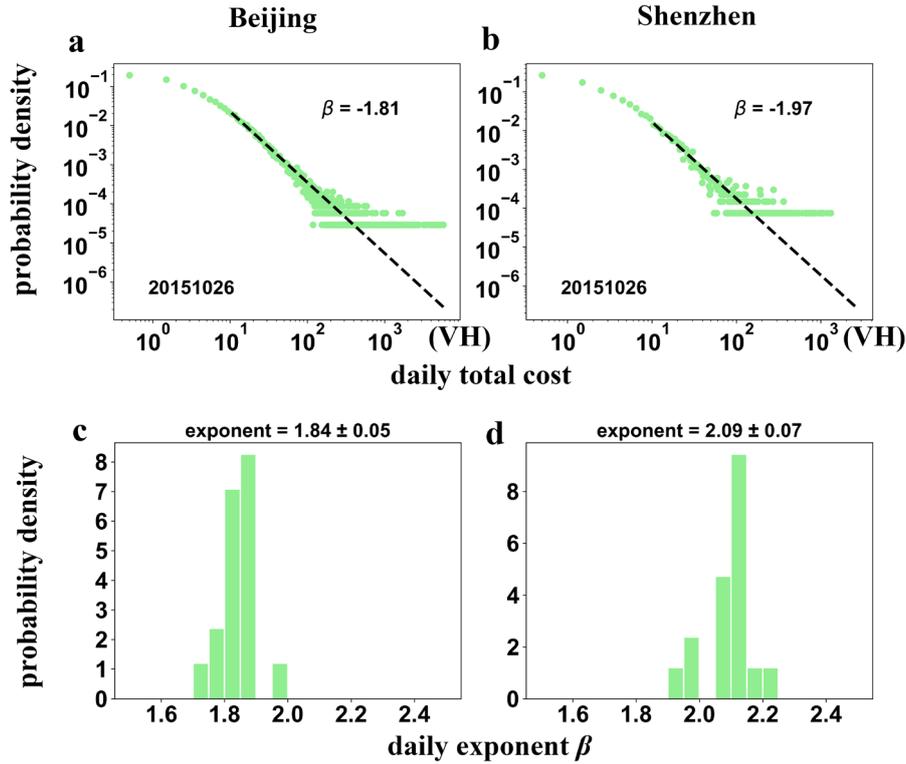

**Fig. 3 Distribution of jam tree costs in urban traffic.** The distribution of jam costs associated with a given trunk in a typical working day for all trunks in **(a)** Beijing and **(b)** Shenzhen. Results for other working days are similar and can be seen in Fig. S2 (Beijing) and Fig. S3 (Shenzhen) in the *Supplementary Information*. Probability density function of the jam cost exponent $\beta$ for 17 working days for **(c)** Beijing and **(d)** Shenzhen. We also apply two-sample Kolmogorov-Smirnov test (KS test) to check if the distributions of daily exponents in two cities are indeed different and obtain a *p*-value of nearly 0 (much lower than the common threshold of 0.05). The distinct values of exponents in these two cities suggests the exponent $\beta$ can represent a "jam-print" of a city.

Next, we focus on the distribution of tree cost of all jam trees during the day. Specifically, for a given day, we sum up the cost of jam trees associated with each specific trunk during the whole day and consider it as the daily cost associated with this trunk. The finding of highly cost trunks is important for identifying, managing, and improving the traffic in these locations which could help improve global urban traffic.

As can be seen in Fig. 3a and Fig. 3b, for a typical working day (Oct. 26, 2015), we find that, for both cities, the probability density of daily cost of jam trees follows a power-law distribution:

$$P(c) \sim c^{-\beta}, \quad (4)$$

where $c$ represents the cost, $P(c)$ is probability density function of the cost distribution, and $\beta$ is the corresponding distribution exponent, with the value of 1.81 and 1.97 for Beijing and Shenzhen respectively in this case. Furthermore, when considering the working days only (over a month), we also find that the exponents of all cost distributions of jam trees for individual days have very similar characteristics with similar exponents. As can be seen in Fig. 3c and Fig. 3d, the cost distribution exponent in Beijing is 1.84±0.05, while in Shenzhen the value is 2.09±0.07. The smaller exponent in Beijing shows that the costs of jams in Beijing are higher compared to Shenzhen. We also calculate the exponents of three other cities including Shanghai, Hangzhou and Jinan, respectively. Results of above cities are shown in Table 2. Based on these results, we argue that these exponents could be considered as a city signature that characterizes the pattern of traffic congestion in a city. This is because the cost distributions are similar on different days for the same city, but rather different in two different cities. Note that smaller distribution exponents represent a more congested city, as they indicate the existence of more large-scale traffic jams in the city.

Table 2: Daily cost distribution exponent $\beta$ of five different cities

| City | Beijing | Shenzhen | Shanghai | Hangzhou | Jinan |
|---|---|---|---|---|---|
| $\beta$ | 1.84±0.05 | 2.09±0.07 | 1.80±0.05 | 1.78±0.07 | 1.75±0.05 |

An important finding of Serok et al.[29] is that the specific bottlenecks of traffic jams, i.e., the jam tree trunks, are poorly predictable, since the repetition of a specific tree trunk of a large jam tree on different days is found to be relatively low, both in London and Tel-Aviv. This finding is also verified here in our large datasets of Beijing and Shenzhen. As can be seen in Fig. S4 in the *Supplementary Information*, it is found that during a week of five working days, the most influential jam trees (i.e., the trunks with cost above 20 VH) that appear repeatably on all five days only account for less than 20% of the total large trees, while over 50% of these trunks reappear in one or two days only. This suggests that most of the heavily costly trunks do not usually recur on a daily basis. Thus, it is difficult to forecast their spatio-temporal location based on historic data. We also calculate the Jaccard Index, often applied to analyze the similarity between two sample sets, to evaluate the overlap of tree trunks between every pair of working days in our dataset. We find that large traffic jams are usually induced by different influential tree trunks on different days (Fig. S5 in the *Supplementary Information*). Particularly, during specific periods (i.e., rush hours or non-rush hours), the overlap of tree trunks with large costs is even lower. Based on our definition of a jam tree, a specific trunk is considered as a specific bottleneck at a fixed location. Therefore, the low result of overlapping trunks on different days suggests that the locations of these bottlenecks, especially the influential ones, are different on different days.

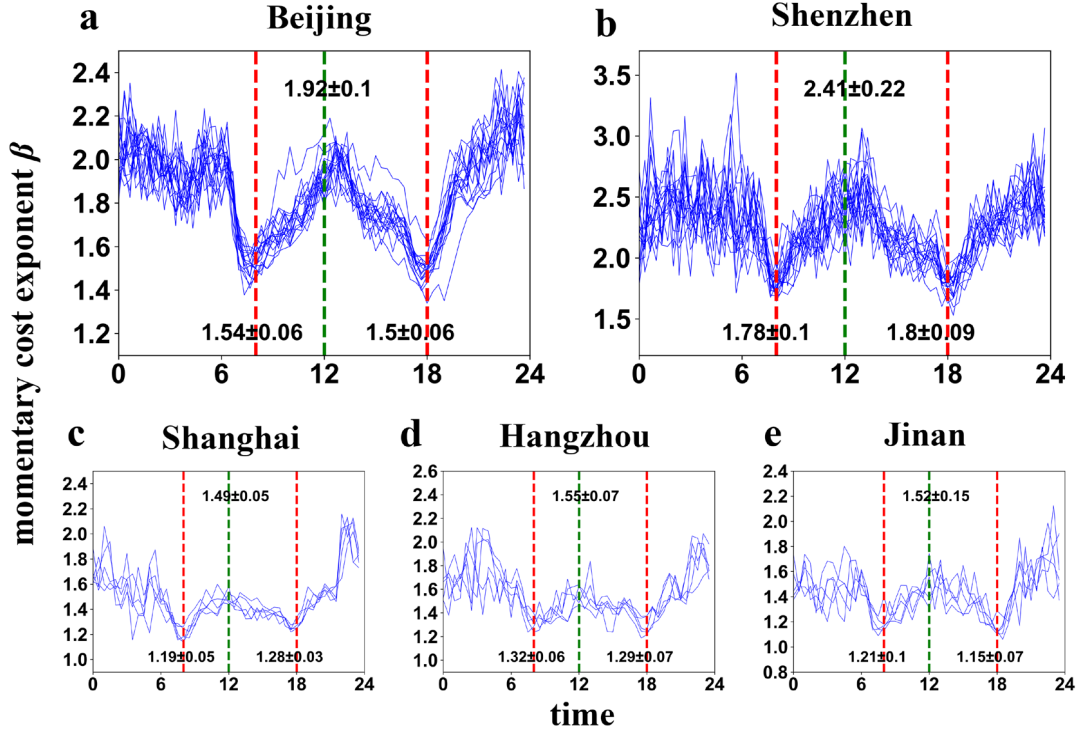

**Fig. 4 Jam patterns of tree cost during the day can be considered as the city jam-prints of urban traffic.** The temporal evolution of the exponent of the cost distribution of jam trees during the day is shown, in 5 different cities **(a)** Beijing (over 17 days), **(b)** Shenzhen (over 17 days), **(c)** Shanghai (over 5 days), **(d)** Hangzhou (over 5 days) and **(e)** Jinan (over 5 days). The numbers in the figures are the mean and standard deviation of daily exponents at three typical time windows (20 min for each) indicated by dash lines, at 8:00AM and 18:00PM for rush hours (red) and at 12:00 for non-rush hours (green).

Based on the above results, we can conclude that while specific traffic congestion, at the microscopic scale, may vary from day to day, the overall macroscopic pattern of traffic congestion in a city (i.e., the distribution of costs throughout the day) remains stable and consistent from day to day. This raises a question of whether there is an indicator that can characterize at the macro-scale the daily evolution patten of traffic congestion in a given region? As the traffic system is expected to present different characteristics during different periods (i.e., rush hours and non-rush hours) the strength of the above indicator, i.e., the exponent of cost distribution $β$, needs to be further validated by testing it at different hours during the day. To this end, we analyze the evolution of exponent $β$ over time during the entire day. We analyze the exponents of the cost distribution of jam trees in 5 different cities in China, during the whole day in different days. For each city, we calculate the tree cost every 20 minutes and analyze the corresponding momentary tree cost distribution and its exponent $β$ by Eq. (4). We find that for each city the evolution of the cost exponent $β$ of the jam trees exhibits a similar temporal "W-like" pattern during working days (as seen in Fig. 4). It is worth noting that the patterns are similar for the same city on different working days but are significantly different for different cities. The "W-like" shape curves for the momentary cost exponent in each day are also consistent with the results found in Fig. 2, indicating that traffic jams are significantly heavier during morning and evening rush hours. Focusing on the evolution of the exponent of the tree cost distribution in different cities, it is interesting to observe that in Shanghai, the traffic situation is typically worse during the morning rush hours than during the evening rush hours. This is represented by smaller momentary cost exponents in the morning rush hour period. However, for the other cities, the momentary cost exponents during the two periods are similar. Overall, the momentary cost exponent for a given time is typically larger in Shenzhen than that in the other cities, suggesting a better traffic flow in Shenzhen compared to other cities.

To test if the congestion pattern of a specific city is unique and characterizes the city, we calculate for different cities, the p-values for the cost distribution exponents during rush hours. As shown in Table 3, by comparing the distribution of the momentary cost during the morning and evening rush hours in each pair of cities (the results in non-rush hours can also be seen in Table S1 in the *Supplementary Information*), we can conclude that generally, different cities have different characteristic values of cost distribution exponents. Although some cities may have similar jam features during a certain period, for instance, morning rush hours in Hangzhou and Jinan or evening rush hours in Shanghai and Hangzhou, we did not find a case that a pair of cities have similar distributions throughout both rush hour periods. These results suggest that the congestion pattern in each city during the day, as shown in Fig. 4, is unique for a city, and can also serve as a jam-print for a city.

**Table 3: The p-values for cost distribution exponents in morning/evening rush hours of pairs of cities**

|  | **Beijing** | **Shenzhen** | **Shanghai** | **Hangzhou** | **Jinan** |
|---|---|---|---|---|---|
| **Beijing** | 1.000 | <0.001 | <0.001 | <0.001 | <0.001 |
| **Shenzhen** | <0.001 | 1.000 | <0.001 | <0.001 | <0.001 |
| **Shanghai** | <0.001 | <0.001 | 1.000 | 0.008 | *0.873* |
| **Hangzhou** | <0.001 | <0.001 | *0.079* | 1.000 | 0.008 |
| **Jinan** | <0.001 | <0.001 | 0.008 | 0.008 | 1.000 |

**Note**: The yellow color represents results comparing the morning rush hours (i.e., 7:00-9:00AM) while green represents results comparing the evening rush hours (i.e., 17:00-19:00PM). If the p-value is lower than threshold of 0.05, the two distributions are regarded as well separated.

Besides single cities, we also consider if the "jam-print" can be generalized to larger scale regions, i.e., the urban agglomeration. Urban agglomeration is a highly developed spatial form of integrated cities where the cooperation among these cities is highly shifted, which renders the region one of the most important carriers for global economic development[31]. Here we focus on the traffic jams on inter- and intra-city highways of a typical urban agglomeration region, i.e., the Beijing-Tianjin-Hebei Urban Agglomeration of China. We first calculate the temporal evolution of jam trees during the workdays, as can be seen in Fig. 5a and Fig. 5b. Similar to single city results, one can observe two typical peaks during the rush hours representing heavier traffic jams during these two periods. However, it is interesting to note that the daily variation of total momentary cost during evening rush hours is much higher than that during morning rush hours. This may indicate that during morning rush hours, traffic flow tends to be more focused or intense, while in the evening traffic jams on highways might occur more sporadically. This could be related to the more dispersed nature of trips in the evening, as people travel to a variety of destinations rather than predominantly to work locations as in the morning.

Next, we analyze the evolution of exponent $β$ during the entire day in urban agglomeration region. We mainly concentrate on jam propagation on the highways in this region. We still calculate the exponent $β$ of momentary tree cost distribution every 20 minutes, and analyze how it evolves over time during the day. As shown in Fig. 5c, the evolution of exponent $β$ of this region also exhibit a similar "W-like" pattern for five different workdays. It can be noticed in this region that the average values of $β$ during morning and evening rush hours are even smaller than any single city aforementioned, with values of approximately 1.11 and 1.08 respectively. This may represent the possible heavier jams in the highways than in the urban central areas. However, during other periods the highways would recover to a more efficient state (e.g., with exponent $β$ around 1.65 at noon). Combination of these different exponents can therefore represent a new jam-print of this urban agglomeration region.

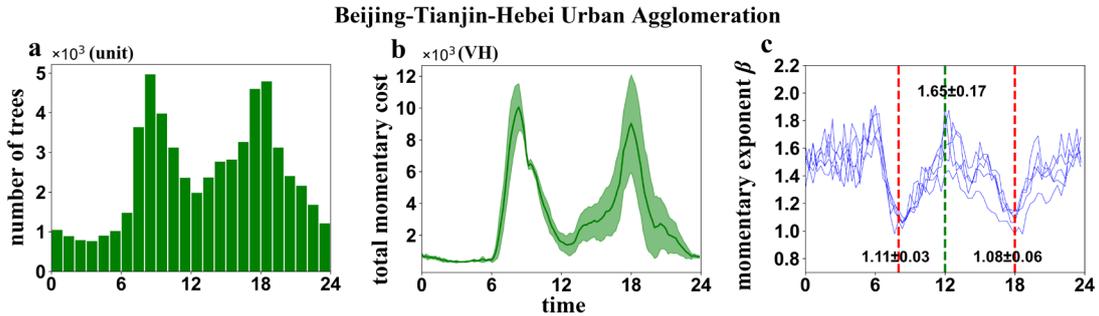

**Fig. 5 Results of jam patterns in highways of an urban agglomeration region.** (a) The number of existing jam trees during the day (averaged over 5 working days with the resolution of 1 hour). (b) The total momentary cost of jam trees evolving with time during the day (with the resolution of 10 minutes). (c) The temporal evolution of the exponent $β$ of the cost distribution of jam trees during the day. The numbers in the figures are the mean and standard deviation of different days at three typical time windows (20 min for each) indicated by dash lines, at 8:00AM and 18:00PM for rush hours (red) and at 12:00 for non-rush hours (green).

**Discussion**

In order to uncover the macroscopic patterns, as well as the spatio-temporal evolution, of traffic congestions in different cities, we study the congestion patterns using real data based on our extended jam tree model which analyzes the propagation of traffic jams associated with specific bottlenecks. The extended jam tree model can detect further configurations of jam trees, compared to Ref. 29. Particularly, it addresses the case where several jam trees share the same trunk or branches, a situation which frequently appears in complex traffic systems (for demo see Fig. 1) such as traffic networks in megacities or even urban agglomerations. Based on the extended model, we study the time evolution of both, the number of existing jam trees and their cost during the day. We also calculate the distribution of the cost of jam trees during a day, and find that the cost distribution of the jam trees follows a power-law distribution based on Eq. (4), with a similar daily exponent for each of the analyzed cities, but different for different cities. Furthermore, we analyze the evolution of the cost distribution exponent during the entire day. We find that the patterns of the cost distribution exponent of the jam trees during the day, can be considered as a jam-print of the urban traffic in a city. This is since the patterns of the evolution of the exponent are similar each day for the same city, but different for different cities. These unique patterns of traffic in urban areas which we suggest as a jam-print of a specific urban traffic may provide valuable insights for determining the quality of traffic in the city and for establishing new traffic management goals and assessing the effectiveness of various traffic improvement strategies.

The consistent power-law scaling observed for the cost distributions of traffic jams across different days in each city may be linked to self-organized criticality[32,33] in the organization of urban traffic. On one hand, an increase in traffic demand during a specific period may lead to the spontaneous appearance of congestion; on the other hand, the individuals seeking the best possible route may redistribute the traffic and drive the system to an optimal state under current traffic conditions. The balance between these two forces could therefore push the traffic system towards its intrinsic operational limits.

We find that each city has its own distinct macroscopic pattern of traffic congestion that is consistent on a daily basis at the same time of day, while the specific locations of traffic jams tend to vary. We hypothesize that the daily variations at the microscopic level might be associated with the fluctuation of human behavior and to the slight daily changes of origin-destination (OD) demand. This phenomenon is analogous to phase transition near criticality, where the distribution of finite clusters consistently follows the same power law while the spatial location of these finite clusters varies significantly in different realizations[34]. Note that power law distributions appear also in other urban features, e.g., in population of cities[35]. Future studies may attempt to focus on uncovering the topological and dynamical mechanisms that could explain the jam-prints of city patterns.

**Methods**
**Data description.** Our data covers five large cities (i.e., Beijing, Shenzhen, Shanghai, Hangzhou and Jinan) and an urban agglomeration (i.e., Beijing-Tianjin-Hebei Urban Agglomeration Region) in China. Our data includes the topology information of road networks and the operational information of traffic dynamics. We obtained information on each road segment which include the road identification information, direction, length, and rank. The identification information of each road segment includes the identification number of its segments, its source intersection, and its target intersection.

The operational information is the real-time velocity records of each road segment with a resolution of 5 minutes up to 1 minute, which is obtained from the global positioning system (GPS) data recorded by floating cars. Specifically, the scales of the road networks and the coverage period of velocity records are shown in Table 4.

Table 4: Key information of the dataset of different cities

| Region | Scale | Duration |
| --- | --- | --- |
| Beijing | road segments: 52968<br>intersections: 27877 | 17 working days in Oct. 2015<br>(resolution of 1min) |
| Shenzhen | road segments: 22248<br>intersections: 12337 | 17 working days in Oct. 2015<br>(resolution of 1min) |
| Shanghai | road segments: 50469<br>intersections: 25729 | 5 working days in Oct. 2015<br>(resolution of 1min) |
| Hangzhou | road segments: 35815<br>intersections: 18591 | 5 working days in Oct. 2015<br>(resolution of 5min) |
| Jinan | road segments: 22690<br>intersections: 12063 | 5 working days in Oct. 2015<br>(resolution of 5min) |
| Beijing-Tianjin-Hebei Urban Agglomeration (highways) | road segments: 66925<br>intersections: 64204 | 5 working days in Mar. 2023<br>(resolution of 1min) |

**Dynamical traffic network.** Based on the topology and operational information of the urban traffic system, we constructed the dynamic traffic network of each region at every moment. For each city, we regard the road segments as network links and the intersections as the network nodes. Then, we constructed the network representation based on their connecting relations. On the basis of the above network representation, we assign a real-time velocity of each road segment to its corresponding link, which can be regarded as the weight of the corresponding link. We used the relative velocity of each road segment as the link weight for normalization. The relative velocity of a given road segment is calculated by dividing its current velocity by its velocity limit. We chose the 95 percentile of the measured velocity during a day as this limit. This is to avoid the influence induced by extreme values. By this, we construct the weighted dynamical traffic network of each region at every moment, where the network topology is static while the network's dynamics is represented by the varying weight of each link.

**Status of the link.** In the dynamic traffic network, we consider the status of a link as either functional or congested based on its average weight within a certain time interval (here 10 minutes). Specifically, we regard the link weight of 0.5 as the relative velocity threshold - a link with an average weight lower than 0.5 is regarded as congested. It is worth noting that the velocity value of a given road segment could be vacant for some moments due to missing data. In this case, we assume that if such a road segment is located between two congested segments, and has the same traffic direction as the congested segments have, it is also highly likely to be congested; otherwise, we will consider the traffic flow on this road segment as fluent. Following this assumption, we can identify the status of every link in the dynamic traffic network for every time interval during the day.

**Trunk and branches of a jam tree.** After identifying the status of each link in the network, we analyzed the properties of jam trees in the dynamic traffic network. The propagation of traffic jams, from downstream roads to upstream roads, is similar to the growth process of a tree. The origin of the traffic jam (which we define as the trunk of the jam tree) is assumed to be the road segment in its downstream with the longest congestion duration compared to other upstream parts of the traffic congestion (which we define as the branches of the jam tree), as shown in Fig. 1. By following this principle, we can identify every jam tree in the dynamical traffic network and locate its trunk and branches by using the following steps:

(i) Calculating the jam duration of each link. The congestion duration of a link is the number of successive time intervals the link has been congested until the current moment.

(ii) Locating the trunk of a jam tree. For a congested link, if there is no neighboring link with a longer congestion duration along its downstream direction, it is assumed to be the trunk of a jam tree.

(iii) Identifying branches belonging to a trunk. Since the traffic congestion propagates from a downstream road to an upstream road, the jam duration of the upstream road should always be shorter than its downstream neighbors which belong to the same jam tree. Based on causality assumption, the difference between two congestion durations of neighboring links must be below a reasonable time threshold $\theta$; otherwise, it can be argued that the congestion of the upstream road may have not been caused by its downstream neighbor. Here we set this threshold $\theta$ as 2 units of time intervals (i.e., 20 minutes)[29]. We identify the nearest neighbor branches of a given jam tree trunk by examining all links in the upstream direction and selecting those that are equal or below the duration threshold criteria to belong to that tree. We proceed to evaluate if any adjacent links connected to the identified branches are also part of the same jam tree, using the duration threshold principle. We repeat this process iteratively, adding new branches to the jam tree until no further branches meet the criteria, and then go back to the first stage, looking for congested links that are not connected to any of the identified jam trees.

**Estimation of number of lanes in each road.** Since in our dataset we do not have information on the number of lanes in each link, we estimate this value based on the rank of each link as in Table 5.

Table 5: Rank and estimated number of lanes for each link (road segment)

| Rank | Road type | Estimated lanes |
| --- | --- | --- |
| 1 or 2 | Highways | 4 |
| 3 | Arterial roads | 3 |
| 4 | Secondary trunk roads | 2 |
| 5 or 6 | Branch way | 1 |

**Data Availability**

The data that support the findings of this study are available on request from the corresponding authors.

**Code Availability**
The code is available upon request directly from corresponding authors.

**Acknowledgments**
This work was supported by the National Natural Science Foundation of China (Grants 72225012, 72288101, 71890973/71890970), the Israel Science Foundation (Grant No. 189/19), the Binational Israel-China Science Foundation (Grant No. 3132/19), Science Minister-Smart Mobility (Grant No. 1001706769), and the European Union's Horizon 2020 research and innovation programme (DIT4Tram, Grant Agreement 953783).

**Author contributions**
**G.Z. and N.S. contributed equally to this work.** G.Z., N.S., E.B.L. and S.H. conceived and designed the research. G.Z., N.S., E.B.L., D.L. and S.H. implemented the methods. G.Z., N.S. and J.D. performed and checked the experiments. All authors analyzed the data. G.Z., N.S., E.B.L, D.L. and S.H. wrote the paper.

**Competing interests**
The authors declare no conflict of interest.

**Additional information**
More figures are presented in **Supplementary Information**.

# Supplementary Information for

# Unveiling City Jam-prints of Urban Traffic based on Jam Patterns

Guanwen Zeng[1,2], Nimrod Serok[3], Efrat Blumenfeld Lieberthal[3], Jinxiao Duan[4], Shiyan Liu[2], Shaobo Sui[2], Daqing Li[2], Shlomo Havlin[1,]*

1. *Department of Physics, Bar-Ilan University, Ramat Gan 52900, Israel.*
2. *School of Reliability and Systems Engineering, Beihang University, Beijing 100191, China.*
3. *Azrieli School of Architecture, Tel Aviv University, Tel Aviv 6997801, Israel.*
4. *School of Economics and Management, Beihang University, Beijing 100191, China.*

**Email**
*Correspondence should be addressed to: havlins@gmail.com

**The SI file includes:**
Supplementary Notes 1 to 5,
Supplementary Figures S1 to S6,
Supplementary Table S1,
and Supplementary References 1.

**Supplementary Note 1: Average momentary cost evolving with time.**

We calculate the average momentary cost evolving with time in 17 working days in Beijing and Shenzhen. By calculating the average momentary cost, at each instant, we sum up the cost of all jam trees throughout 17 days and divide it by the total number of trees, as shown in Fig. S1.

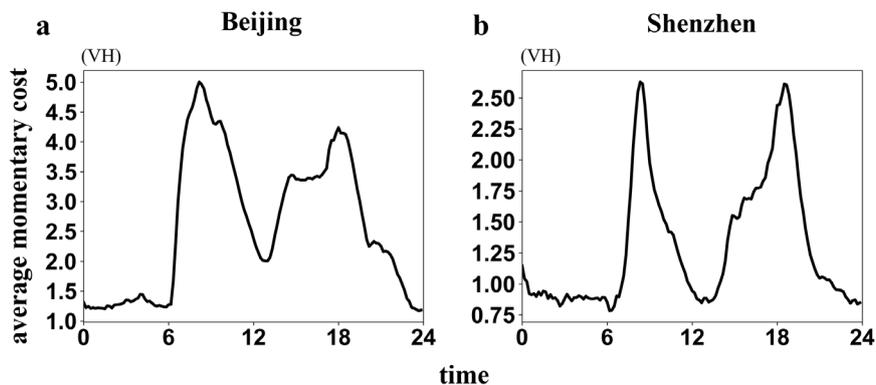

**Fig. S1** The average momentary cost in 17 working days evolving with time in **(a)** Beijing and **(b)** Shenzhen

**Supplementary Note 2: Distribution of daily jam costs and its corresponding exponents.**

We calculate the distribution of jam costs associated with a given trunk in each working day in Beijing and Shenzhen. Moreover, we also calculate the probability density of the jam cost exponent $\beta$ by Eq. (4) for 17 working days to check if the daily cost exponent is stable. Note that in Ref. 1, it is found that the distribution exponents of the average daily total tree cost for three typical urban areas, ranges from 1.6 to 1.7, which is lower than the results of Beijing and Shenzhen, see Fig. S2 and Fig. S3.

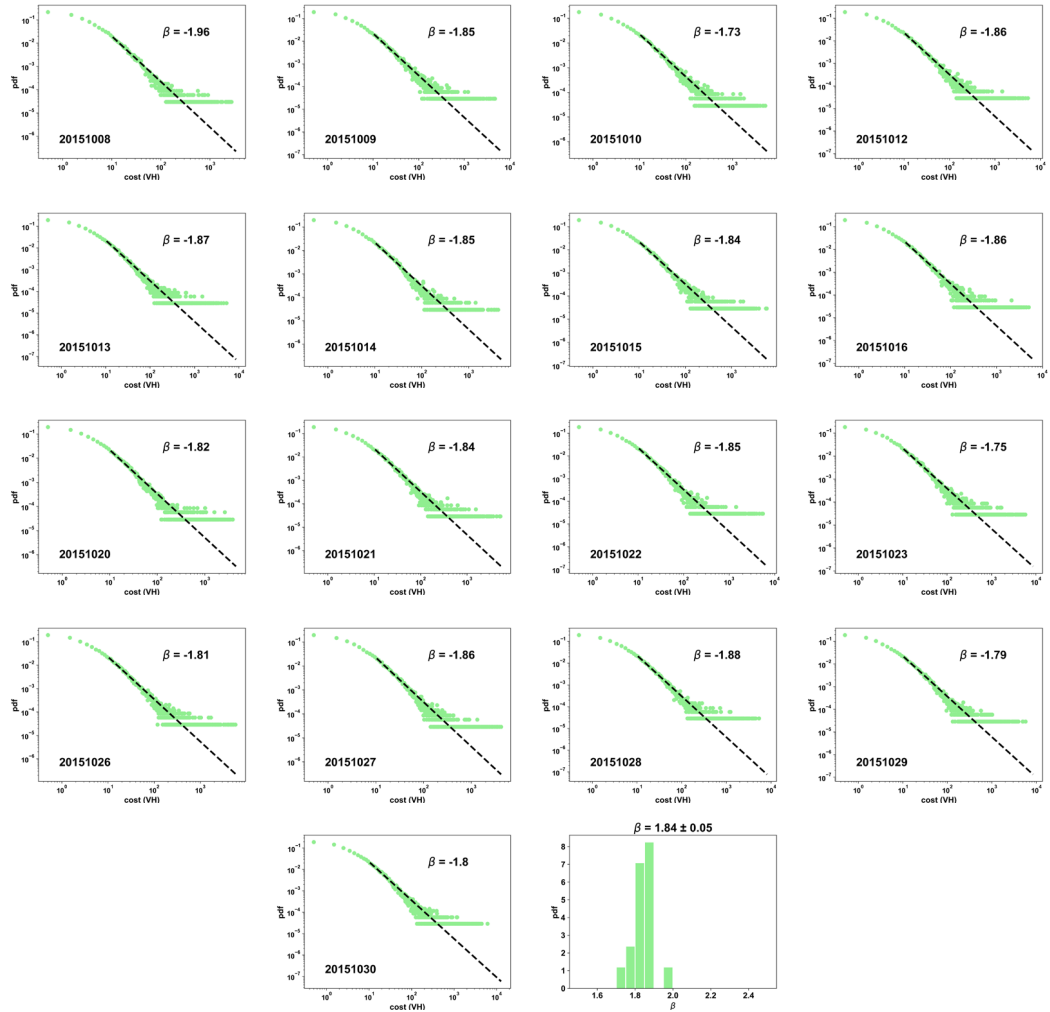

**Fig. S2** Daily total cost distribution of jam trees for Beijing, fitting the exponent $\beta$ for cost $\geq 10$ (VH). The value of the exponent $\beta$ is shown for each day. The probability density of the daily exponents is shown in the last subfigure.

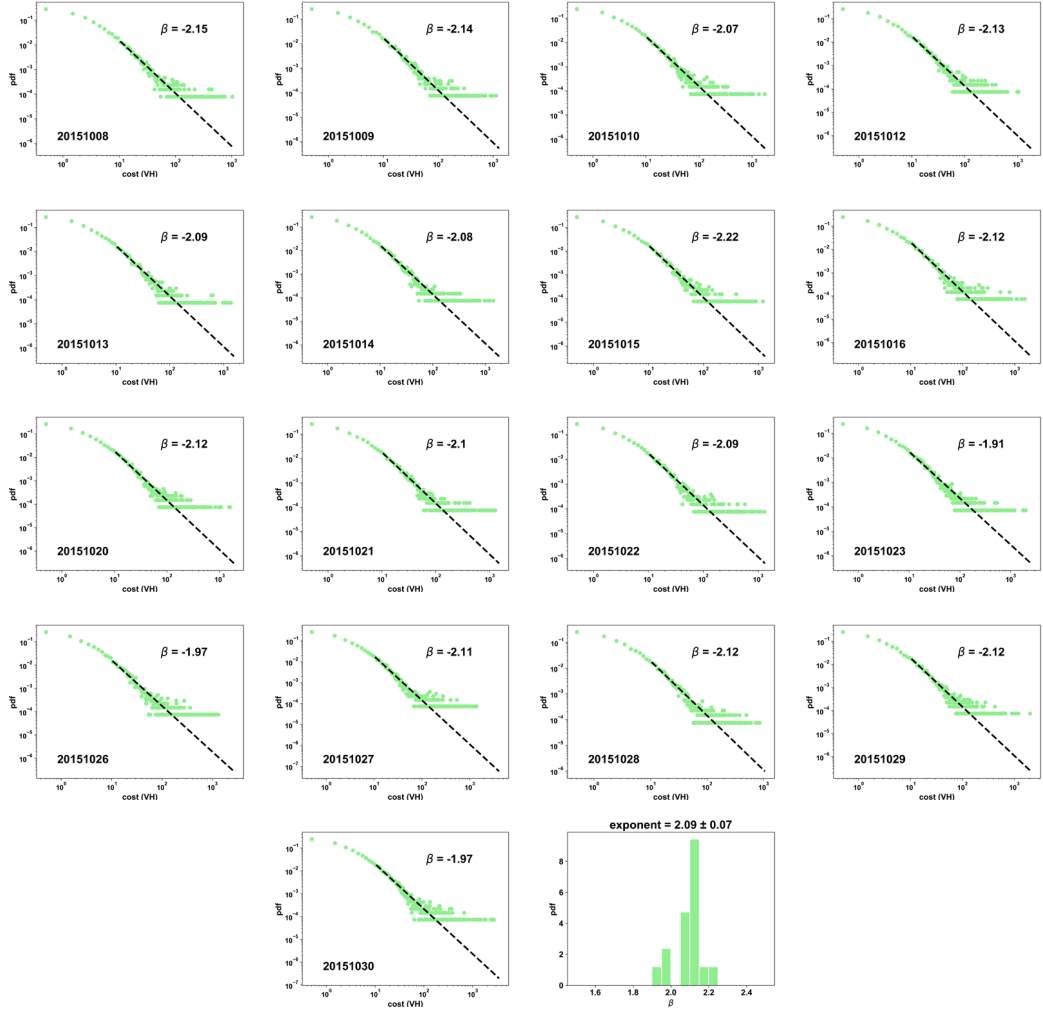

**Fig. S3** Daily total cost distribution of jam trees for Shenzhen, fitting the exponent $\beta$ for cost ≥ 10 (VH). The value of the exponent $\beta$ is shown for each day. The probability density of the daily exponents is shown in the last subfigure.

**Supplementary Note 3: Low recurrence of specific trunks.**

We find in our large datasets of Beijing and Shenzhen that most of the heavily costly trunks do not usually recur on a daily basis, which is consistent with the findings in Ref.[S1]. This verifies the argument that specific bottlenecks of traffic jams, i.e., the jam tree trunks, are poorly predictable. This conclusion is supported by calculating both the repetition (see Fig. S4) and the Jaccard index (see Fig. S5) of specific trunks.

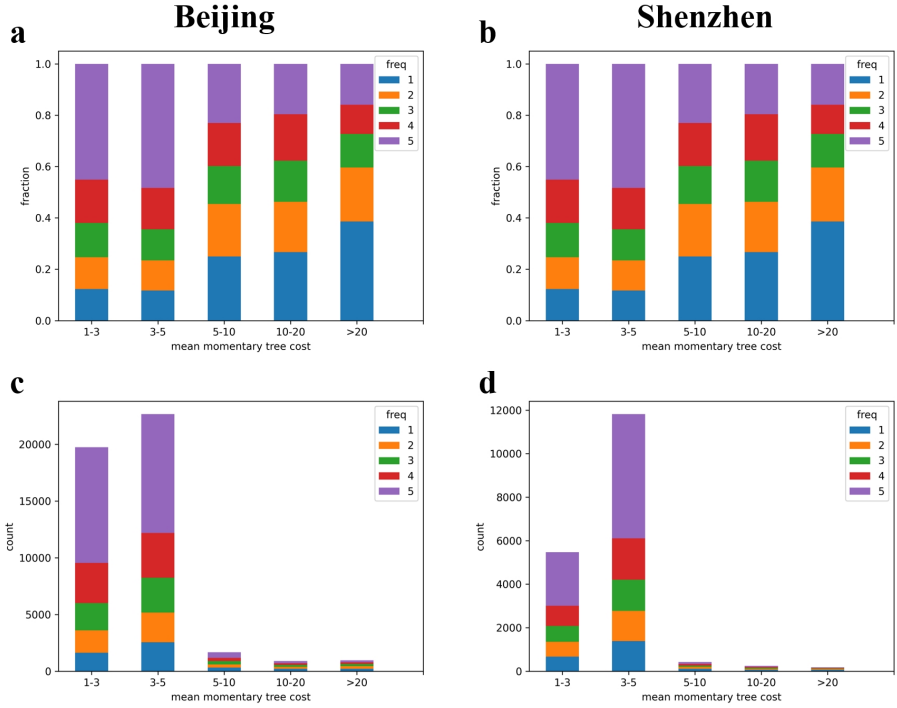

**Fig. S4** Repetition for specific trunks. **(a)** and **(c)** Show the results of Beijing while **(b)** and **(d)** show the results of Shenzhen. We first calculate the mean momentary tree cost for each specific trunk over five workdays (i.e., from Oct. 26 to Oct. 30, 2015), and classify all the trunks into different categories based on this cost (as in X-axis). Next, we calculate, for each category, the fraction and count of the trunks that repeated during the measured days.

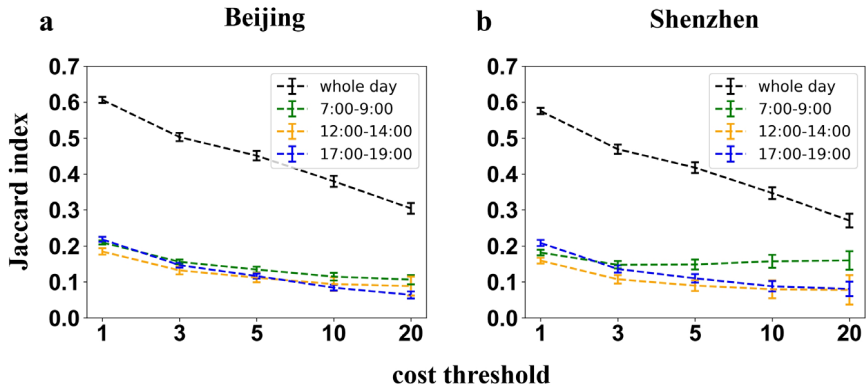

**Fig. S5** The Jaccard index for specific trunks. The Jaccard index tests the overlap of specific trunks in different days in **(a)** Beijing and **(b)** Shenzhen. For each day, we used a vector of trunks $\mathbf{v} = [c_1, c_2, \ldots, c_n]$ to represent the daily state, where $c_1, c_2, \ldots, c_n$ are the mean momentary tree cost of each specific trunk. Under the given cost threshold $c$, we selected the trunk with tree cost $> c$ to construct a new vector for each day and calculate the Jaccard index (mean value and error bar) between different days.

**Supplementary Note 4: Distribution of cost exponents during different periods.**

We also calculate the distribution of cost exponents for more specific periods, i.e., rush hours and non-rush hours, as shown in Fig. S6.

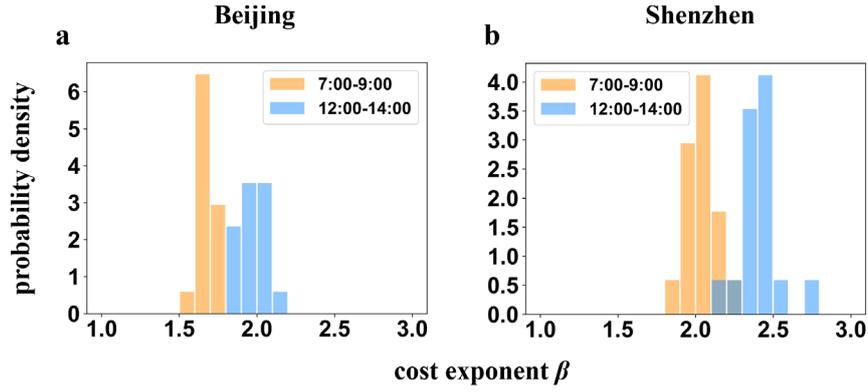

**Fig. S6** Distribution of cost exponents for different days during rush hours and non-rush hours periods for **(a)** Beijing and **(b)** Shenzhen. The orange bars represent the rush hour periods between 7:00-9:00AM while the blue bars represent the non-rush hour periods between 12:00-14:00PM for each city. Note the lower values of the exponents for Beijing compared to Shenzhen mean larger traffic jam trees in Beijing.

**Supplementary Note 5: Statistical test for congestion patterns during non-rush hours.**

We compare here the distributions of the momentary cost during the non-rush hours, i.e., 12:00-14:00PM, for different cities. As seen in Table S1, during non-rush hours, Beijing and Shenzhen have different jam patterns compared to other cities, while Shanghai, Hangzhou and Jinan are similar in jam patterns.

Table S1: p-values for cost distribution exponents in non-rush hours (12:00-14:00PM)

|  | Beijing | Shenzhen | Shanghai | Hangzhou | Jinan |
|---|---|---|---|---|---|
| Beijing | 1.000 | <0.001 | <0.001 | <0.001 | <0.001 |
| Shenzhen | / | 1.000 | <0.001 | <0.001 | <0.001 |
| Shanghai | / | / | 1.000 | *0.079* | *0.357* |
| Hangzhou | / | / | / | 1.000 | *0.357* |
| Jinan | / | / | / | / | 1.000 |

**Note**: If the p-value is lower than threshold of 0.05, the two distributions are regarded as well separated.